\documentclass[11pt]{article}
\usepackage{amsmath}
\usepackage{amssymb}
\usepackage{cite}
\usepackage[cp1251]{inputenc}
\usepackage{amsfonts}
\usepackage{graphicx}
\usepackage{latexsym}
\usepackage{amssymb}
\usepackage{amsmath}
\usepackage{amsfonts}
\usepackage{mathrsfs}
\usepackage{cite}
\usepackage{bm}
\usepackage{color}
\usepackage[english]{babel}
\usepackage{indentfirst}

\newcommand{\p}[1]{(\ref{#1})}
\def\nn{\nonumber\\}
\def\eq{\eqref}

\def\nb{\nabla}
\def\nn{\nonumber}
\def\cN{{\cal N}}
\def\cD{{\cal D}}
\def\cA{{\cal A}}

\def\cV{{\cal V}}
\def\cW{{\cal W}}

\def\bea{\begin{eqnarray}}
\def\tr{{\rm Tr }}
\def\beq{\begin{gather}}
\def\eea{\end{eqnarray}}
\def\eeq{\end{gather}}
\def\ln{\,\mbox{ln}\,}

\def\Det{\,\mbox{Det}\,}

\def\tr{\,\mbox{tr}\,}

\def\Tr{\,\mbox{Tr}\,}

\def\al{\alpha}
\def\da{\dot{\alpha}}

\def\f{\frac}
\def\sB{\stackrel{\frown}{\square}}
\def\bsB{\stackrel{\frown}{\bm{\square}}}

\begin{document}

\begin{titlepage}

\begin{center}
\vspace{1cm} {\Large\bf On a structure of the one-loop divergences
in $4D$ harmonic superspace sigma-model}

\vspace{1.5cm} {\bf I.L. Buchbinder\footnote{joseph@tspu.edu.ru
}$^{\,a,b,c}$, A.S.
Budekhina\footnote{budekhina@tspu.edu.ru}$^{\,a,b}$, B.S.
Merzlikin\footnote{merzlikin@tspu.edu.ru}$^{\,a,c}$ }

\vspace{0.4cm}

{\it $^a$ Center for Theoretical Physics, Tomsk State Pedagogical
University, \\
634061, Tomsk, Russia
\\ \vskip
0.15cm $^b$ National Research Tomsk State University, 634050, Tomsk,
Russia
\\ \vskip
0.15cm
$^c$  Tomsk State University of Control Systems and Radioelectronics,\\
634050, Tomsk, Russia\\
}
\end{center}
\vspace{0.4cm}
\begin{abstract}
We study the quantum structure of four-dimensional $\cN=2$
superfield sigma-model formulated in harmonic superspace in terms of
the omega-hypermultiplet superfield $\omega$. The model is described
by harmonic superfield sigma-model metric $g_{ab}(\omega)$ and two
potential-like superfields $L^{++}_{a}(\omega)$ and
$L^{(+4)}(\omega)$. In bosonic component sector this model describes
some hyper-K\"{a}hler manifold. The manifestly $\cN=2$
supersymmetric covariant background-quantum splitting is constructed
and the superfield proper-time technique is developed to calculate
the one-loop effective action. The one-loop divergences of the
superfield effective action are found for arbitrary $g_{ab}(\omega),
L^{++}_{a}(\omega), L^{(+4)}(\omega)$, where some specific analogy
between the algebra of covariant derivatives in the sigma-model and
the corresponding algebra in the $\cN=2$ SYM theory is used. The
component structure of divergences in the bosonic sector is
discussed.
\end{abstract}
\end{titlepage}

\setcounter{footnote}{0} \setcounter{page}{1}

\section{Introduction}
Nonlinear sigma-models have been studied extensively over the years
and take a special place in many areas of the field theory. One
attractive class of such models is given by the two-dimensional
conformal field theories, which are well known to be exactly
solvable in the sense that S-matrix, the correlation functions and
the anomalous dimensions are completely determined by the conformal
invariance \cite{BPZ}. Such  models have non-trivial UV behaviour
being finite in the case of Ricci-flat target spaces
\cite{AGF1,AFM}. Besides, in the worldsheet field theory, the
correlations between conformal invariance, (extended) supersymmetry
and geometry of the complex manifolds in the full quantum theory
lead to restrictions on the background fields geometry in every
order of the perturbation theory \cite{AGF,BW1,ACG,ACG1}.

Supersymmetric two-dimensional nonlinear sigma-models caused the
great interest due to their remarkable properties and inspired the
construction of the geometrical non-polynomial theories of the
supersymmetric matter in four-dimensional space-time (see, e.g.
\cite{BKet,DK,Ketov0}). The formulation of supersymmetric nonlinear
sigma-models has been worked out both for the simple and extended
supersymmetries (see, e.g., \cite{GGRS,BK1,GIOS} for review). The
examination of these theories led to the discovery of fascinating
relations between extended supersymmetry and complex manifolds. It
was shown that the supersymmetric sigma-models in four dimensions
possess the K\"{a}hler manifolds as a target
\cite{Zum,HKLR,GIOS-85,BW} in case of $\cN=1$ supersymmetry and the
hyper-K\"{a}hler manifolds for the $\cN=2$ case
\cite{AGF,ST-83,GIOS-85,GIOS-88} (see also \cite{Kuz12} for a
review). A number of supersymmetric sigma-models and their chiral
truncations, constructed within ten-dimensional superstring theory
\cite{GSW}, where the extra dimensions are wrapped up into a coset
space, has been of a certain interest. Also, note the construction
of the one-dimensional $\cN=4$ sigma-model in harmonic superspace
\cite{FIS}.

The present paper studies the quantum aspects of the
four-dimensional $\cN=2$ supersymmetric sigma-model in harmonic
superspace \cite{GIOS}. We introduce the harmonic superfield
sigma-model given in terms of analytic superfield
$\omega$-hypermultiplet. As well known, there are two types of
hypermultiplets in harmonic superspace, the $q$-hypermultiplet and
$\omega$-hypermultiplet (see, e.g. \cite{GIOS}). Formulated in terms
of unconstrained $\cN = 2 $ superfields, such a model, as well as
the general $q$-hypermultiplet theory (see, e.g. \cite{GIOS}), has
two manifest symmetries: reparametrization invariance in the
harmonic superspace and $\cN=2$ supersymmetry. However, use of
$\omega$-hypermultiplet leads to certain simplifications in
constructing the quantum effective action, since
$\omega$-hypermultiplet is a uncharged superfield. Here we would
like to make a comment on terminology. According to \cite{GIOS-88}
(see also \cite{GIOS}) the $4D$ hyper-K\"{a}hler sigma-model is
associated with general $q$-hypermultiplet theory in harmonic
superspace.  Such a general theory is formulated in terms of
functions $L^{+}_a(q^{+},u)$ and $L^{+4}(q^{+},u)$ of hypermultiplet
analytic superfields $q^{+a}$ and harmonics $u$. The theory is
manifestly invariant under the $\cN=2$ supersymmetry and arbitrary
reparameterizatios of the hypermultipet superfield. After going to
components and eliminating the auxiliary fields, we get in bosonic
sector an action of the $4D$ hyper-K\"{a}hler sigma-model. In our
paper we introduce a model which is the superfield sigma-model with
harmonic superfield target space and harmonic superspace metric.
Such a model is of course related of general $q$-hypermultiplet
theory. To avoid ambiguities, we call the considered theory the
harmonic superfield sigma-model. This theory is interesting by
itself, since it is internally consistent and similar in many
aspects of the conventional sigma model due to the presence
superfield metrics in harmonic superspace.  As well as the general
$q$-hypermultipet theory, the harmonic superspace sigma-model is
manifestly invariant under the $\cN=2$ supersymmetry and arbitrary
reparameterizatios of the hypermultipet superfield. The classical
aspects of $\cN = 2$ hypermultiplet models in various dimensions are
widely studied (see, e.g., \cite{GIOS,KM,BP1} for a review). The
quantum structure of harmonic superspace sigma-models has never been
studied in detail. The aim of this paper is to fill this gap.

The main object of our work is the quantum effective action of the
harmonic superfield  sigma-model. Since the classical model under
consideration possesses two manifest symmetries, it is natural to
develop such a scheme for constructing the effective action that
preserves the same manifest symmetries. As a result, we are faced
with the problem of a manifest covariant formulation of the
effective action and the problem of its manifest covariant
computation. The solution to the first problem is realized within
the harmonic superspace background field method based on superfield
background-quantum splitting that generalizes the known
background-quantum splitting procedure in the conventional
sigma-models (see,
e.g.,\cite{Spence,AGF,ACG,ACG1,HPS,Mukhi,BDS})\footnote{Quantization
of the general $q$-hypermultiplet model faces a problem of
preserving the classical reparameterization invariance in quantum
theory since a standard loop expansion destroys this symmetry. The
advantage of harmonic superspace sigma-model is that it possesses a
natural geometric structure in harmonic superspace and therefore we
can use the covariant loop expansion analogous to conventional
sigma-model.}. The second problem is solved with the help of
superfield proper-time technique, which is a powerful tool for
manifest covariant analysis of the effective actions in the
supersymmetric field theories (see the applications of this
technique in the various superfield models e.g., in
\cite{BK1,BBIKO,BBKO,BKT,BBIKO1,KM1,KM2,Kuzenko,Kuzenko2,KM1}). As
we will see, just use of the $\omega$-hypermultiplet allows to use
simply enough the superfield proper-time technique. Note also that
the harmonic superspace approach is the only manifestly $\cN = 2$
supersymmetric formalism that may preserve the explicit off-shell
supersymmetry on all steps of quantum computations (see, e.g.,
\cite{GGRS,BBP}).

The divergences of the effective action in four-dimensional $\cN=1$
supersymmetric sigma-models are studied in \cite{Spence} in the case
of vanishing (anti-)chiral potentials and in \cite{BBP} in the
general case. Some assumptions about a structure of the possible
one-loop divergences in $\cN=2$ sigma-models on the base of $\cN=1$
divergences were considered in \cite{Spence}. In the present paper,
we calculate the one-loop divergent contributions to the effective
action of $\cN=2$ supersymmetric sigma-model in manifestly covariant
and $\cN=2$ supersymmetric manner, which as far as we know, was not
held directly in terms of $\cN=2$ superfields. It is worth pointing
out that the calculation of divergences in $\cN=2$ supersymmetric
sigma-models has a common difficulty: the absence of analytic normal
coordinates on the generic K\"{a}hler manifolds does not allow to
develop a higher-loop expansion preserving all symmetries of the
theory. This fact has already been  mentioned in the pioneering
papers \cite{AGF1,AFM}. Some papers were directly addressed to the
treatment of the above difficulty \cite{CL,HN1,HN2}. However, this
problem is unessential for one-loop calculations because the above
difficulty arises only in the higher loops.

The paper is organized as follows. In Section 2, we discuss the
basic properties of $\cN=2$ harmonic superspace and formulate the
harmonic superfield sigma-model.  In Section 3, we develop the
covariant background field method to study the one-loop effective
action of the model. For these aims, we consider the covariant
harmonic and spinor derivatives and study their algebra. Section 4
is devoted to the calculation of the one-loop divergent
contributions of the effective action. Section 5 includes the
discussion of the main obtained results and the directions of
further studies. In the Appendix we discuss how the harmonic
superfield sigma-model is connected to general q-hypermultiplet
theory.

\section{Harmonic superfield sigma-model}

Throughout the paper, we use the notations and conventions from
\cite{GIOS}. Namely, we denote by
 \bea
(z,u)=(x^M,\theta^{\al}_i,\bar{\theta}^{i}_{\da},u^{\pm i}),\qquad
M=0,..,3, \quad \alpha =1,2 \quad i=1,2,
 \eea
the central basis coordinates of the $\cN=2$ harmonic superspace.
The additional harmonic variables $u^{\pm i}$ correspond to the
coset of the R-symmetry group $SU(2)/U(1)$ of the $\cN=2$ Poincare
superalgebra in four dimensions. The analytic harmonic superspace
involves the coordinates
 \bea
(\zeta,u) = (x^M_{\cal A},
\theta^{+}_{\al},\bar{\theta}^{+}_{\da},u^{\pm i}), \qquad x^M_{\cal
A}=x^M - 2i  \theta^{(i}\sigma^M \bar{\theta}^{j)}u^{+}_{ i}u^{-}_{
j}, \quad  \theta^{+ \al}= u^+_i\theta^{\al i}.
 \eea
The analytic harmonic superspace involves only half of the original
Grassmann coordinates and is closed under the $\cN=2$ supersymmetry
superspace \cite{GIOS}.

Let us introduce the spinor and harmonic derivatives in the analytic
basis
 \bea \label{DeriVat} D^{+}_\alpha &=&\frac{\partial}{\partial
\theta^{-\alpha}}, \quad
  \bar{D}^{+}_{\da}=\frac{\partial}{\partial\bar{\theta}^{-\da}},\\
  D^{-}_{\alpha}&=&-\frac{\partial}{\partial \theta^{+\alpha}}
  +2i\bar{\theta}^{-\da}\partial_{\alpha\da}, \quad
\bar{D}^{-}_{\da}=-\frac{\partial}{\partial \bar{\theta}^{+\da}}+2i
\theta^{-\alpha}\partial_{\alpha\da},\label{DeriVat1}
  \eea
where $\partial_{\alpha\da} = (\sigma^{M})_{\al\da} \partial_{M}$,
$(\sigma^{0})_{\al\da}$ is the unit matrix and
$(\sigma^{i})_{\al\da}$ are the Pauli matrices. The spinor
derivatives \eqref{DeriVat} and  \eqref{DeriVat1} together with
harmonic derivatives
  \bea
 D^{\pm\pm}&=&u^{+i}\frac{\partial}{\partial u^{-i}}
 -2i\theta^{+\alpha}\bar{\theta}^{+\da}\partial_{\alpha\da}
            +\theta^{+\alpha}\frac{\partial}{\partial \theta^{-\alpha}}
            +\bar{\theta}^{+\da}\frac{\partial}{\partial \bar{\theta}^{-\da}},\\
D^0 &=& u^{+i}\frac{\partial}{\partial u^{+i}} -
u^{-i}\frac{\partial}{\partial u^{-i}} +
    \theta^{+\alpha}\frac{\partial}{\partial \theta^{+\alpha}}
    +\bar{\theta}^{+\da}\frac{\partial}{\partial \bar{\theta}^{+\da}} -
    \theta^{-\alpha}
    \frac{\partial}{\partial \theta^{-\alpha}}
     -\bar{\theta}^{-\da}\frac{\partial}{\partial \bar{\theta}^{-\da}},
 \eea
satisfy the algebra
 \bea
  &&[D^{++}, D^{--}] = D^0\,,\quad  [D^{\pm\pm},D^\pm_{\al}]=0\,,
  \quad  [D^{\pm\pm},D^\pm_{\da}]=0\,,\nonumber\\
   && [D^{\pm\pm},D^\mp_{\al}]=D^{\pm}_{\al}\,,
   \quad \ [D^{\pm\pm},D^\mp_{\da}]=D^{\pm}_{\da}\,,
   \quad \{\bar{D}^{+}_{\da},D^{-}_{\al}\}=-\{D^{+}_{\al},\bar{D}^{-}_{\da}\}=2i\partial_{\al\da}\,.
\label{alg1}
 \eea

The following conventions for the full and the analytic superspace
integration measures are used
  \bea
d^{12} z  = d^4 x_{\cal A} (D^{+})^4 (D^{-})^4, \quad d^{8} z  = d^4
x_{\cal A} (D^{+})^2 (D^{-})^2 , \quad d\zeta^{(-4)} = d^4 x_{\cal
A} (D^{-})^4 du,
 \eea
where we have assumed the notation
   \bea
 &&(D^{+})^4=(D^{+})^2(\bar D^{+})^2, \qquad (D^{-})^4=(D^{-})^2(\bar D^{-})^2,\\
 && (D^{\pm})^2=\frac{1}{4}D^{\pm\alpha}D^{\pm}_\alpha,
  \qquad (\bar{D}^{\pm})^2=\frac{1}{4}\bar{D}^{\pm}_{\da}\bar{D}^{\pm\da}.
  \eea

We consider a four-dimensional $\cN=2$ supersymmetric sigma-model in
the analytic harmonic superspace. The model is described in terms of
analytic harmonic superfields $\omega^a (\zeta,u),$ which
parameterize the $n$-dimensional target space, $a=1,..,n$. The
classical action for the model is written in the form
   \bea\label{classaction}
  S[\omega]=\int d\zeta^{(-4)}\left(
  -\frac{1}{2}g_{ab}(\omega) D^{++}\omega^{a}D^{++}\omega^{b}
  +L^{++}_{a}(\omega)D^{++}\omega^{a}+L^{(+4)}(\omega)\right),
   \eea
where the target space metric $g_{ab}$, and  $L^{++}_{a}$ and
$L^{(+4)}$ are the arbitrary analytic functions of the
$\omega^{a}$-superfields\footnote{Since the omega-hypermultiplet is
uncharged, the functions $L^{++}_{a}$ and $L^{(+4)}$ must obligatory
include the explicit dependence on harmonics.}. The action
\eqref{classaction} is invariant under reparameterizations
transformations
 \bea
\omega^{a} \rightarrow \omega^{a} + \lambda^{a}(\omega,u)\,,
\label{trans}
      \eea
in the assumption that superfields $g_{ab},\, L^{++}_a$ and
$L^{(+4)}$ transform under \eqref{trans} as a  tensor  of the
corresponding rank. The equations of the motion corresponding to
action (\ref{classaction}) looks like
 \bea
(D^{++})^2\omega^{a}
 + \Gamma^{a}{}_{bc}(\omega)D^{++}\omega^{b}D^{++}\omega^{c}
 + g^{ab}L^{++}_{bc}(\omega)D^{++}\omega^{c} + g^{ab}\partial_{b}L^{+4}(\omega)=0,
 \label{eqm}
 \eea
where we have introduced the harmonic superspace Christoffel symbols
$\Gamma^{a}{}_{bc}(\omega)$ and the inverse metric $g^{ab}$. Also we
have denoted
$L^{++}_{ab}=\partial_{a}L^{++}_{b}-\partial_{b}L^{++}_{a}.$ The
superfield model with action (\ref{classaction}) as well as
conventional sigma-model contains the sigma-model-type metric and
two derivatives. Therefore, the study of quantum structure of this
model can be based on a generalization of the methods developed for
conventional sigma models.

\section{Background-quantum splitting}

In this section, we develop the covariant background field method in
$\cN=2$ harmonic superspace to study  the effective action for the
model \eqref{classaction}. It is known that the linear
background-quantum splitting to construct the loop expansion  of the
quantum effective action for the nonlinear sigma-models leads to
breaking the reparametrization invariance. To preserve the above
invariance in quantum theory for nonlinear sigma-models, the
manifestly covariant background field formalism was developed (see,
e.g. \cite{Mukhi,HPS}). Here we adopt such a formalism for the
$\omega$-hypermultiplet in harmonic superspace.

Let us introduce the analytic superfield $\rho^{a}(s)$ that
satisfies the harmonic superspace geodesic equation
 \bea
 \label{geoeq}
\frac{d^2 \rho^{a}(s)}{ds^2}+\Gamma^{a}{}_{bc}(\rho)\frac{d
\rho^{b}(s)}{ds}\frac{d\rho^{c}(s)}{ds}=0,
 \eea
with the conditions
 \bea
\rho^a(0)=\Omega^a, \quad \rho^a(1) =\Omega^a + \pi^a, \quad \f{d
\rho^a}{d s}\Big|_{s=0} = \xi^a\,. \label{geocond}
 \eea
The analytic superfield $\xi^a = \xi^a(\zeta,u)$ denotes the tangent
vector to the geodesic at $s=0$ and plays the role of quantum field
in the background-quantum splitting.

The solution to the equation \eqref{geoeq} with the initial
conditions \eq{geocond} can be written in the form of the series
 \bea
 \rho^a(s) = \Omega^{a} + \sum_{n=1}^\infty\frac{s^n}{n!}
 \rho^{a}_{(n)}\,,
 \label{decomp}
 \eea
where the functions $\rho^{s}_{(n)}$ depend on the background fields
$\Omega^a$ and quantum $\xi^{a}$ ones and are obtained directly from
the equation \eq{geoeq}
 \bea
\rho^{a}_{(1)} = \xi^a, \quad \rho^{a}_{(2)} =
-\Gamma^a{}_{bc}\xi^b\xi^c, \quad \rho^{a}_{(3)} = - (\partial_d
\Gamma^a{}_{bc} - 2\Gamma^a{}_{bk}\Gamma^k{}_{cd})\xi^b\xi^c\xi^d,
\qquad {\rm ect.}
 \eea

The decomposition of the classical action \p{classaction} under
\eq{decomp}
 \bea
S[\rho]=S[\Omega]+\sum_{n=1}^{\infty}\frac{1}{n!}\frac{d^n
S[\rho]}{ds^n}\Big|_{s=0} = S[\Omega] + S_1 + S_2 + \dots.
\label{decomp2}
 \eea
will be manifestly covariant. Our aim is to study the one-loop
quantum correction to the classical action \eqref{classaction}. For
this aim, only the quadratic in quantum fields $\xi^a$ part in
action $S_2$ in \eq{decomp2} should be taken into account.  The
explicit expression for $S_2$ is written as follows
 \bea
 S_2 &=&\frac{1}{2}\int
 d\zeta^{-4}\,\xi^a\Big(g_{cd}(\nabla^{++})^c_a (\nabla^{++})^d_b
 -R^{d}{}_{abc}\,D^{++}\Omega^{c} D^{++}\Omega_{d}
 +\nabla_{(a}L^{++}_{c}(\nabla^{++})^{c}_{b)}
\nonumber \\
&& + \nabla_{a}\nabla_{b}L^{++}_{c}D^{++}\Omega^{c}
+L^{++}_{d}R^{d}{}_{abc}D^{++}\Omega^{c}
+\nabla_{a}\nabla_{b}L^{(+4)}\Big)\xi^{b}, \label{S2}
 \eea
where we have introduced the harmonic covariant derivative
 \bea
 \label{der}
 (\nabla^{++}\xi)^a=(\nabla^{++})^a{}_b \xi^{b} = D^{++}\xi^a
 + \Gamma^{a}{}_{bc}(\Omega)D^{++}\Omega^{c}\xi^{b},
 \eea
and $\nabla_{a}$ is a covariant derivative along  the  curve
$\rho^a(s)$ in the target space. We also assume the symmetrization
without one half in the expression \eq{S2}. Riemann tensor
$R^{d}{}_{abc}$ and the metric tensor $g_{ab}$, which appeared above
in \eq{S2}, depend on the background superfields $\Omega^a$. The
following useful property
 \bea
(\nabla^{++}g)_{ab} =0,
 \eea
can be derived from the vanishing of the covariant derivative of the
metric tensor.


\section{One-loop divergences}

The quadratic over quantum fields $\xi^{a}$ action $S_2$, explicitly
derived in \eqref{S2}, determines the one-loop quantum correction
$\Gamma^{(1)}$ to the classical action \eqref{classaction}. After
integrating over quantum fields $\xi^a$, one obtains
 \bea
\Gamma^{(1)}[\Omega] = \Det^{-\frac{1}{2}} (S''_2[\Omega]) =
\tfrac{i}2 \Tr_{(4,0)} \ln S''_2[\Omega]\,, \label{G}
 \eea
where the functional trace includes the matrix trace and integration
over harmonic superspace
$$
\Tr_{(q,4-q)} {\cal O} = \tr \int d \zeta_1^{(-4)}d \zeta_2^{(-4)}\,
\delta_{\cal A}^{(4-q,q)}(2|1)\,  {\cal O}^{(q,4-q)}(1|2).
$$
The delta-function $ \delta_{\cal A}^{(4-q,q)}(2|1)$ introduced
above is an analytic function on both arguments \cite{GIOS}. Also we
have denoted by ${\cal O}^{(q,4-q)}$ the kernel of the operator
${\cal O}$ acting in the space of analytic superfields with the
harmonic U(1) charge $q$.

The expression \eqref{G} is expressed through the second variation
derivative of the action $S_2$ \eqref{S2} with respect to the
quantum superfields and can be represented in the schematic form
 \bea
\Gamma^{(1)}[\Omega] = \tfrac{i}2 \Tr_{(4,0)} \ln
\Big((\nabla^{++})^2 + (\nabla L^{++}) \nabla^{++}+X^{(+4)}\Big)\,,
\label{G0}
 \eea
where the operator $(\nabla L^{++}) \nabla^{++}$ means $
\label{first derivarive}
\nabla_{(a}L^{++}_{c}(\nabla^{++})^{c}{}_{b)}. $ The expression
$X^{(+4)}$ in \eq{G0} is written as follows
 \bea
X^{(+4)}_{ab}&=& - R^{d}{}_{abc}D^{++}\Omega^{c}D^{++}\Omega_{d} +
\nabla_{a}\nabla_{b}L^{++}_{c}D^{++}\Omega^{c}
\nonumber \\
&& + L^{++}_{d}R^{d}{}_{abc}D^{++}\Omega^{c} +
\nabla_{a}\nabla_{b}L^{(+4)}, \label{Psi}
 \eea
and is symmetrized over the target space indices without one-half.

For further analysis, it is convenient to rewrite the operator
$(\nabla^{++})^2 + (\nabla L^{++}) \nabla^{++}+X^{(+4)}$ in
expression \eq{G0} in another form, where the term is linear in the
operator $\nabla^{++}$ is eliminated with help of redefinition of
covariant derivative. It can be done if to introduce the new
covariant derivative ${\cal D}^{++}=  D^{++} + {\cal V}^{++}  =
\nabla^{++} + \tilde \Gamma^{++}$ in terms of new analytic
connection ${\cal V}^{++}=\Gamma^{++} + \tilde \Gamma^{++}$. Here
$(\Gamma^{++})^a{}_b = \Gamma^a{}_{bc}(\Omega) D^{++}\Omega^c$ and
$(\tilde \Gamma^{++})^a{}_b = g^{ac}\nabla_{c} L^{++}_{b}$. As a
result we obtain
 \bea
 \Gamma^{(1)} = \tfrac{i}{2} \Tr_{(4,0)} \ln \Big( ({\cal D}^{++})^2
 + \tilde X^{(+4)}
 \Big).
 \label{G10}
 \eea
The superfield $\tilde X^{(+4)}$ in \eq{G10} reads
 \bea
 \tilde X^{(+4)}_{ab}&=&-R^{d}{}_{abc}D^{++}\Omega^{c}D^{++}\Omega_{d}
 + L^{++}_{d}R^{d}{}_{abc}D^{++}\Omega^{c}
\nonumber \\
&&  - \nb^c L^{++}_a\, \nb_c L^{++}_b - (\nabla^{++})^c_a \nb_c
L^{++}_b +\nabla_{a}\nabla_{b}L^{++}_{c}D^{++}\Omega^{c} +
\nabla_{a}\nabla_{b}L^{(+4)}. \label{tildePsi}
 \eea
The term linear in covariant derivative $\cD^{++}$ is absent in
expression (\ref{G10}).

The covariant derivative ${\cal D}^{++}=D^{++} + {\cal V}^{++}$ is
similar to the covariant derivative for the hypermultiplet coupled
to gauge superfield, where the connection ${\cal V}^{++}$ is
analogous to gauge superconnection.  The only difference is that in
our case, the connection ${\cal V}^{++}$ is constructed from the
background hypermultiplet while the gauge superconnection is
independent analytic superfield. Note that such kind of connection
was observed and discussed for the first time in the paper
\cite{GIOS-85}\footnote{The connection ${\cal V}^{++}$ in the case
under consideration contains two terms, the harmonic superfield
Christoffel symbols and the term including the $L_{a}^{++}$. In the
case of flat metric $g_{ab}(\Omega)$ we get the connection
introduced in \cite{GIOS-85}.}. Taking into account such an analogy,
the calculations of the effective action (\ref{G10}) can be carried
out using the methods developed earlier for the study of the
effective action in ${\cal N}=2$ supersymmetric quantum gauge theory
(see, e.g. \cite{BBIKO,BBKO,BKT,BBIKO1,KM1,KM2}).

Evaluation of the effective action in ${\cal N}=2$ supersymmetric
quantum gauge theory is based on the algebra of the covariant
derivative. Therefore we begin the calculation of effective action
with a discussion how the initial algebra of supersymmetry
\eqref{alg1} is deformed due to covariantization of the harmonic
derivative $D^{++}$. First of all the covariant derivative ${\cal
D}^{++}$ preserves the analyticity and hence the commutators
$[D^{+}_{\al},{\cal D}^{++}]=[\bar D^{+}_{\da},{\cal D}^{++}]=0$
should be held. Then we define the nonanalytic covariant harmonic
derivative ${\cal D}^{--}$. Similar to the $\cN=2$ SYM theory in
harmonic superspace \cite{GIOS}, we use the zero curvature condition
 \bea
 \label{ncc}
 \big[({\cal D}^{++})^{a}_{c},({\cal D}^{--})^{c}_{b}\big]=\delta^{a}_{b}D_{0}.
 \eea
Assuming ${\cal D}^{--} = D^{--} + {\cal V}^{--}$ we obtain like in
$\cN=2$ SYM theory \cite{GIOS}
 \bea
{\cal V}^{--} = \sum^{\infty}_{n=1}(-1)^{n}\int du_1 \dots
du_{n}\frac{{\cal V}^{++}_1 {\cal V}^{++}_2 \dots {\cal
V}^{++}_n}{(u^{+}u^{+}_{1})\dots(u^{+}_{n}u^{+}) }. \eea

In the rest, the algebra of covariant derivatives looks similar to
the algebra  of $\cN=2$ supersymmetric gauge theory in four
dimensions \cite{GIOS}. Omitting the target space indices, one gets
 \bea
&&
 \{D^{+}_{\al},\cD^{-}_{\beta}\}=2\varepsilon_{\al\beta}\bar{\cal W},
 \qquad \{\bar{D}^{+}_{\da},\bar{\cD}^{-}_{\dot{\beta}}\}=2\varepsilon_{\da\dot{\beta}}{\cal W}, \nn \\
&&
 \{\bar{D}^{+}_{\da},\cD^{-}_{\al}\}= -\{D^{+}_{\al},\bar{\cD}^{-}_{\da}\}=2i\cD_{\al\da}, \nn \\
&&
 \big[D^{+}_{\al},\cD_{\beta\dot{\beta}}\big]
=\bar{D}^{+}_{\dot{\beta}}\varepsilon_{\al\beta}\bar{\cW}, \qquad
\big[\cD^{-}_{\al},\cD_{\beta\dot{\beta}}\big]
=\bar{\cD}^{-}_{\dot{\beta}}\varepsilon_{\al\beta}\bar{\cW},\nn \\
&&
 \big[\bar{D}^{+}_{\da},\nabla_{\beta\dot{\beta}}\big]
=D^{+}_{\beta}\varepsilon_{\da\dot{\beta}}\cW, \qquad
\big[\bar{\cD}^{-}_{\da},\cD_{\beta\dot{\beta}}\big]
=\cD^{-}_{\beta}\varepsilon_{\da\dot{\beta}}\cW\nn \\
&&
 [\cD^{++},\bar \cD^-_{\da}]=\bar D^{+}_{\da}, \qquad
[\cD^{--},D^+_{\al}]=\cD^{- }_{\al}\,. \label{algebra}
 \eea
Here we have denoted
 \bea
 && \cD^-_{\al}=D^-_\al - D^+_\al\cV^{--},\qquad
\cD_{\al\da}=\partial_{\al\da}-\tfrac{i}{2}D^{+}_{\al}\bar{D}^{+}_{\da}\cV^{--},\nn \\
&& \bar{\cW}=(D^{+})^2\cV^{--},\quad \cW=(\bar{D}^{+})^2\cV^{--}.
\label{def}
 \eea

Using the above algebra, we introduce the analytic covariant
d'Alembertian
 \bea
\bsB&=&\tfrac{1}{2}(D^+)^4(\cD^{--})^2 \nn \\
 &=&\cD_{M}\cD^{M}
 -\tfrac{1}{4}({D}^+)^2{\cW}\,\cD^{--}
 -\tfrac{1}{2}D^{+}_{\al}\cW\cD^{-\al}
-\tfrac{1}{2}\bar{D}^{+}_{\da}\bar{\cW}\bar{\cD}^{-\da}\nonumber
\\ && +\tfrac{1}{8}\cD^{-}_{\al}D^{+\al}\cW - \tfrac{1}{2}\cW\bar{\cW}.
\label{sB}
 \eea
The covariant d'Alembertian $\bsB$  depends on the background
$\Omega$ superfield through the analytic connection ${\cal V}^{++}$.
Acting in the space of analytic superfields, this operator preserves
analyticity, $[D^+_\al, \bsB]=0$.

To calculate the one-loop divergent contributions to the effective
action \eq{G10}, we represent it as a sum of two terms
  \bea
 \Gamma^{(1)}[\Omega] = i \Tr_{(2,2)}\ln {\cal D}^{++}
+\tfrac{i}2\Tr_{(4,0)} \ln\Big(\delta^{(0,4)} + G^{(0,0)}\tilde
X^{+4} \Big), \label{G1}
 \eea
where the Green function $G^{(0,0)}$ satisfies the equation
 \bea
({\cal D}^{++}_1)^2 G^{(0,0)}(1,2)=\delta^{(4,0)}_{\cal A}(1,2).
 \eea
Explicit solution this equation has the form \cite{GIOS}
 \bea
G^{(0,0)}(1,2)=-\frac{1}{\bsB_1}(D_1^{+})^4(D_2^{+})^4\delta^{12}(z_1-z_2)\frac{(u_1^-
u_2^-)}{(u_1^+ u_2^+)^3}, \label{Gr}
 \eea
where $\delta^{12} (z_1-z_2)$ is a full $\cN=2$ superspace
delta-function.  It should be noted that the operator $({\cal
D}^{++})^2$ commute with the covariant d'Alembertian  $\bsB$ acting
on the analytic superfields with the zero harmonic $U(1)$ charge.
Dependence on background hypermultiplet superfields $\Omega^a$ is
contained in the operator $\bsB$.

\subsection{Calculating the divergences}

The first term in \eq{G1} is a trace of first order operator acting
in the space of analytic superfields with the harmonic $U(1)$ charge
$+2$. To extract the divergent contribution from this term, we
vary\footnote{Here and after, we follow a similar consideration in
case of the four-dimensional $\cN=2$ supersymmetric gauge theory,
which was carried out thoroughly using the harmonic superspace
formulation in the works \cite{BBIKO,KM2}.} it with respect to the
superfield ${\cal V}^{++}$
 \bea
\delta \Gamma_{\cal W}^{(1)}[\Omega] = i \delta  \Tr_{(2,2)}\ln
{\cal D}^{++} = i\Tr\delta {\cal V}^{++} G^{(1,1)} = i\tr \int
d\zeta_1^{(-4)}\delta{\cal V}^{++}_1 \, G^{(1,1)}(1|2)\Big|_{2=1}\,,
\label{vG}
 \eea
where the trace has been taken over target space indices and
$G^{(1,1)}$ is the Green function of the operator ${\cal D}^{++}$
  \bea
  \cD^{++} G^{(1,1)}(1|2) &=& \delta^{(3,1)}_{\cA}(1|2)\,,\nn \\
G^{(1,1)}(1,2)&=&-\frac{1}{\bsB_1}(D_1^{+})^4(D_2^{+})^4\frac{\delta^{12}(z_1-z_2)}{(u_1^+
u_2^+)^3}, \label{Gr2}
 \eea
For further analysis, we use in \eqref{vG} the proper-time
representation for the operator $\bsB{}^{-1}$ in the Green function
\eq{Gr2}. It leads to
 \bea
\delta \Gamma_{\cal W}^{(1)} =-i\tr\int d\zeta_{1}^{(-4)}\delta
{\cal V}_1^{++} \int_0^\infty d(is)(is\mu^2)^{\f\varepsilon2}
e^{-is\bsB_1}(D_1^+)^4(D^+_2)^4
\f{\delta^{12}(z_1-z_2)}{(u^+_1u^+_2)^3}\bigg|_{2=1}. \label{G4}
 \eea
Here we have introduced the integration over proper-time $s$ and
$\mu$ and $\varepsilon$ are the parameters related to dimensional
regularization. The  divergences of effective action appear  in
\eqref{G4} as a pole $\tfrac1\varepsilon$.

To extract the divergent contributions from the expression \eq{G4},
we first of all, count a coincident points limit,
$\theta_2\to\theta_1$. The integrand in the expression \eq{G4}
contains eight $D$-factors acting on the Grassmann delta-function
$\delta^8(\theta_1-\theta_2)$. Using the identity
 \begin{equation}
(D_1^+)^4(D^+_2)^4 \delta^8(\theta_1-\theta_2)\big|_{2\to 1} =
(u^+_1u^+_2)^4 \label{id}
 \end{equation}
one rewrites the expression \eq{G4} in the form
 \bea
\delta \Gamma^{(1)}_{{\cal W},\,\rm div}= -i\tr\int
d\zeta_{1}^{(-4)}\delta{\cal V}_1^{++} \int_0^\infty
d(is)(is\mu^2)^{\f\varepsilon2} e^{-is\sB_1}(u^+_1u^+_2)
\delta^{4}(x_1-x_2)\Big|_{2=1, \,\rm div}. \label{G5}
 \eea

Then we commute the operator $e^{-is \bsB}$ with the factor
$(u_1^+u_2^+)$, taking into account that the operator $\bsB$
\eqref{sB} contains the harmonic derivative $\cD^{--}$. In the
coincident harmonic limit, $u_2^+\to u_1^+$, the only non-zero
contribution appears due to the identity
$\cD_1^{--}(u^+_1u^+_2)|_{2\to 1}  = (u^-_1u^+_2)|_{2\to 1} =-1$.
After that one obtains
 \bea
 e^{-is\bsB}(u^+_1u^+_2)e^{is\bsB}\Big|_{2=1} = \tfrac{s}{4}({D}^+)^2{\cW} + \dots
 \eea
where we omit all terms with high power of the proper-time $s$. Such
contributions correspond to the finite corrections to the effective
action. Next, we pass to the momentum representation for the
space-time delta-function and calculate the integral over
proper-time. It leads to
 \bea
\delta \Gamma^{(1)}_{\cW,\, \rm div} = \frac{1}{(4\pi)^2\varepsilon}
\tr\int d\zeta^{-4} \delta\cV^{++} ({D}^{+})^2{\cW}. \label{G6}
 \eea
Hence, from eq. \eq{G6} one can read off the divergent part of
$\Gamma_{\cal W}^{(1)}$
 \bea
\Gamma^{(1)}_{\cW,\, \rm div} = \frac{1}{2(4\pi)^2\varepsilon}
\tr\int d^8z \,\cW^2, \label{G7}
 \eea
where the superfield $\cW$ was defined in \eqref{def}.

Now let us consider the second term in the effective action
\eqref{G1}. This part of effective action is defined as a series
 \bea
\Gamma^{(1)}_{X} = \tfrac{i}2\Tr_{(4,0)} \ln\Big({ \bf 1} +
(G^{(0,0)})^{ac}\tilde
X^{(+4)}_{cb}\Big)=\tfrac{i}2\Tr\sum_{n=1}^{\infty} \frac{(-1)^n}{n}
\Big((G^{(0,0)})^{ac}\tilde X^{(+4)}_{cb}\Big)^n\,, \label{G7}
 \eea
where the matrix power in the last expression includes the
integration over analytic subspace as well. The Green function
$G^{(0,0)}$ contains the inverse power of the d'Alembertian $\bsB$.
By power counting the divergent contributions correspond to the term
with $n=2$ in the decomposition \eq{G7}
 \bea
\Gamma^{(1)}_{X,\,\rm div}&=&\frac{i}{4}\int
d\zeta_1^{(-4)}d\zeta_2^{(-4)}\, \,(G^{(0,0)})^{ab}(1|2)\, \tilde
X^{(+4)}_{bc}(1) (G^{(0,0)})^{cd}(2|1)\, \tilde X^{(+4)}_{da}(2).
 \eea

The further strategy is as follows. The Green function $G^{(0,0)}$
\eq{G0} contains the $D$-factors, which can be used to restore the
full superspace measure. Then, we integrate over Grassmann variable
$\theta_2$, taking into account the delta-function
$\delta^{8}(\theta_2-\theta_1)$ from the Green function $G^{(0,0)}$,
and use the identity \eq{id}. After that, one gets
 \bea
\Gamma^{(1)}_{X,\,\rm div}=-\frac{1}{4(4\pi)^2 \varepsilon}\int
d^{12}z \,du_1 du_2\, \frac{(u_1^{-}u_2^{-})^2}{(u_1^{+}u_2^{+})^2}
\tilde X^{(+4)ab}(1) \tilde X^{(+4)}_{ba}(2), \label{G8}
 \eea
where we have used the divergent part of the momentum integral
 \bea
 \int \frac{d^4 q}{q^2(q-p)^2} =  \frac{i\pi^2}{\varepsilon}, \quad \varepsilon \to 0\,.
 \eea

Finally,  we combine both contributions \eq{G6} and \eq{G8}, and
obtain the divergent contribution of the effective action \eq{G1} in
the form
 \bea
\Gamma^{(1)}_{\rm div}&=&\frac{1}{2(4\pi)^2\varepsilon} \tr\int d^8z \, {\cal W}^2\nonumber \\
&& -\frac{1}{4(4\pi)^2\varepsilon}\int d^{12}z\, du_1
du_2\frac{(u_1^{-}u_2^{-})^2}{(u_1^{+}u_2^{+})^2} \, \tilde
X^{(+4)ab}(1) \, \tilde X^{(+4)}_{ba}(2), \label{Psi2}
 \eea
The relation (\ref{Psi2}) is the general result for the divergences
of the one-loop effective action (\ref{G0}). The function $\tilde
X^{(+4)ab}$ is given by (\ref{tildePsi}). It is interesting to note
two points. First, the expression (\ref{Psi2}) is invariant under
the reparameterizations transformations (\ref{trans}) by
construction. Second, the divergent part of effective action is
non-local in harmonics. This should not come as much of a surprise.
For example, the classical action of $\cN=2$ SYM theory is non-local
in harmonics as well \cite{GIOS}. However, the corresponding
component action is completely local. Later we will show that the
expressions of the form (\ref{Psi2}) are also local in the
components.

\subsection{Special cases}

Let us consider the expression \eq{Psi2} in more detail.  The
superfield $\tilde X^{(+4)}$ was introduced in \eq{tildePsi} and
included two types of contributions. The first one depends on the
Riemann tensor, constructed in the background metric tensor
$g_{ab}(\Omega)$. Such terms vanish in the case of the constant
metric, $g_{ab}[\Omega] = h_{ab}$. The second type of contributions
contains both the background metric and the potential functions
$L^{++}_a$ and $L^{(+4)}$. Such type of contributions vanish if we
assume $L^{++}_a=0 $ and $L^{(+4)}=0$. Using the expression
\eq{Psi2}, one can extract them independently.

First, let us assume $L^{++}_a=0 $ and $L^{(+4)}=0$ and mark the
corresponding effective action by symbol R. In this case the
one-loop contribution to effective action \eq{G1} is reduced to
 \bea
\Gamma^{(1)}_{R}[\Omega]=i \Tr_{(2,2)} \ln \nabla^{++}
+\tfrac{i}2\Tr_{(4,0)} \ln\Big({ \bf 1} -
(G^{(0,0)})^{ac}R^{d}{}_{cbe}D^{++}\Omega^{e}D^{++}\Omega_{d} \Big),
\label{w}
 \eea
where the covariant derivative $\nabla^{++}$ was introduced in
\eq{der}. The divergent contribution, in this case, can be found
from the general relation (\ref{Psi2}) and has the form
 \bea
\Gamma^{(1)}_{R,\,\rm div}[\Omega] &=& \frac{1}{2(4\pi)^2\varepsilon} \tr\int d^8z \, W^2 \nn \\
&& -\frac{1}{4(4\pi)^2 \varepsilon}\int d^{12}z \,du_1 du_2\,
\frac{(u_1^{-}u_2^{-})^2}{(u_1^{+}u_2^{+})^2}
R^{c a b}{}_{ d}(1)R^{e}{}_{b a k}(2) \nn \\
&& \qquad \qquad \times D^{++}\Omega^{d}(1) D^{++}\Omega_{c}(1)
D^{++}\Omega^{k}(2) D^{++}\Omega_{e}(2). \label{G9}
 \eea
The superfield $W$ in the expression \eq{G9} is obtained from $\cW$
\eq{def} in case of vanishing potential functions $L_a^{++}$ and
$L^{(+4)}$. The superfield connection $\cV^{++}$ in such a case
coincides with the Levi-Civita analytic connection
$(\Gamma^{++})^a_b = \Gamma^a{}_{bc}(\Omega) D^{++}\Omega^c$.

Now we return to the effective action \eq{G1} and consider the case,
when the background metric, $g_{ab}(\Omega)= h_{ab}$, does not
depend on the superfield $\Omega$ and superspace point $z$. In this
case, all terms containing Christoffel symbols and Riemann curvature
tensor obviously disappear, and we have only parts depending on
potentials $L^{++}_{a}, L^{(+4)}$. We mark the corresponding
effective action by symbol L. In this case, the one-loop effective
action \p{G1} is reduced to
 \bea
\Gamma^{(1)}_{L}[\Omega] &=& i \Tr_{(2,2)} \ln \tilde{\nabla}^{++} +
\tfrac{i}2\Tr_{(4,0)} \ln\Big({ \bf 1} + G^{(0,0)} {L}^{(+4)} \Big),
\label{Gflat}
 \eea
where we have introduced the notations
  \bea
{\nabla}^{++}_{ab}&=&h_{ab}D^{++}+\tilde \Gamma^{++}_{ab},\\
{L}^{(+4)}_{ab}&=&\partial_{a}\partial_{b}L^{(+4)}
+\partial_{a}\partial_{b}L^{++}_{c}D^{++}\Omega^{c}-D^{++}\tilde{\Gamma}^{++}_{ba}
+ (\tilde{\Gamma}^{++})^2_{ab}. \label{var}
 \eea
The analytic superfield $\tilde \Gamma^{++}_{ab}
=\partial_{a}L^{++}_{b}$ can be considered as an analytic connection
$\cV^{++}$ corresponding to the case under consideration. The
superfield $L^{++}_{ab}$ was introduced earlier in \eq{eqm}. The
divergent part of the effective action \eqref{Gflat} is obtained
from general relation (\ref{Psi2}) and has the form
 \bea
\Gamma^{(1)}_{L,\,\rm div}&=&\frac{1}{2(4\pi)^2\varepsilon} \tr\int d^8z \, \tilde{W}^2\nonumber \\
&& -\frac{1}{4(4\pi)^2\varepsilon}\int d^{12}z\, du_1
du_2\frac{(u_1^{-}u_2^{-})^2}{(u_1^{+}u_2^{+})^2} \, {L}^{(+4)}_{ab}
(1) \, {L}^{(+4)}{}^{ba}(2). \label{L2}
 \eea
The superfield $\tilde W$ is defined by the superfield $\cW$
\eq{def} for the case of the constant metric $g_{ab}$, when
superfield connection $\cV^{++}$ reducing to  $\tilde
\Gamma^{++}_{ab}$. The structure of the contribution above is
similar to \eq{Psi2} and \eq{G9}.

\subsection{The component structure of divergences}

Let us briefly discuss the component structure of \eqref{G9} and
\eq{L2}. We are not going to consider the whole component form of
the one-loop divergences and will only demonstrate that the
non-local in harmonics expressions of the form (\ref{Psi2}) are
local in components. To see that it is sufficient to derive the
component form of the expression (\ref{w}) in bosonic sector since
it completely reflects the general expression (\ref{Psi2}).

The component expansion of the background field $\Omega^a
(\zeta,u),$ contains both the physical fields and the set of
auxiliary fields. Using the classical equations of motion \eq{eqm},
one can exclude auxiliary unphysical components and formulate the
model in terms of the physical fields (see the \cite{GIOS} for
details). The divergent contributions \eqref{G9} and \eq{L2} to the
one-loop effective action were calculated without any restriction on
the background superfields $\Omega^a (\zeta,u)$. Hence, we can use
the general expression for the decomposition of the analytic
superfield $\Omega$ in series over Grassmann variables $\theta^+$
and $\bar \theta^+$
 \bea
\Omega (\zeta,u) = \tfrac{1}{\sqrt{2}}\omega
(x)+\omega^{ij}(x)u^{+}_{i}u^{-}_{j}+\theta^{+\al}\psi^{i}_{\al}(x)u^{-}_{i}+
\bar{\theta}^{+}_{\da}\bar{\psi}^{\da}_{i}(x)u^{-i}+
\dots,\label{component}
 \eea
where ellipses stand for the unessential (for us) terms. The fields
$\omega$ and $\omega^{ij} = \omega^{ji}$ are the physical bosonic
scalar fields, and $\psi^{i}_{\al}(x) $ is the spinor one. In the
expression \eq{component}, we are assuming that all fields have the
target space indices.

The general scheme of passing to component action in the divergent
contribution \eq{G9} is as follows. First, we substitute the
component expansion \p{component} for the superfield $\Omega$ and
collect the terms with the fourth power of Grassman variables
$\theta$ and $\bar{\theta}$. Then we use the relations $u^{+}_i
u^{-}_j - u^{+}_j u^{-}_i=\varepsilon_{ij},\, u^{+i} u^-_i =1$, to
separate the explicit dependence of the integrand in \eqref{G9} on
the harmonic variable $u_2$ and integrate over $u_2$ using  the
general rules of integration over harmonics \cite{GIOS}. After that
the integrals over anticommuting  variables $\theta$ and
$\bar\theta$ are evaluated and for the result in the bosonic sector
of the expression \eq{G9} we have
 \bea
\Gamma^{(1)}_{\rm div}[\omega] =-\frac{1}{128\pi^2\varepsilon} \int
d^4x\, R^{c a b}{}_{ d}R^{e}{}_{b a k} \,
\partial_{\al\da}\omega^{d}\, \partial^{\al\da}\, \omega_c
\partial_{\beta\dot{\beta}}\omega^{k} \partial^{\beta\dot{\beta}}\omega_{e}+\dots,
\label{component action}
 \eea
where we omit all terms with permutations on spinor indices and
terms with the triplet of scalars of bosonic component $\omega_{ij}$
of the superfield $\Omega$. The expression (60) is evidently local.

The component expression of the divergent contribution \p{L2} can be
obtained in the same manner. In this case,  the explicit form of the
component action depends on the potential-like functions
$L^{++}_{a}$ and $L^{(+4)}$.

\section{Summary}

In the present paper we have developed the manifestly covariant
approach for studying the quantum structure of the harmonic
superspace sigma-model in four dimensions. Specific feature of this
approach is the exploration of the analogies between one-loop
effective action for such a theory and one-loop effective action for
hypermultiplet in external non-Abelian $\cN=2$ supersymmetric gauge
superfield.

The harmonic superspace sigma-model (\ref{classaction}), is
formulated in $\cN=2$ harmonic superspace in terms of analytic
omega-hypermultiplet superfields. Such a formulation provides both
manifest reparameterization invariance in harmonic superspace and
manifest $\cN=2$ supersymmetry. The one-loop effective action for
such a model is constructed in the framework of the manifestly
covariant and manifestly $\cN=2$ supersymmetric background-quantum
splitting in $\cN=2$ harmonic superspace. Such an one-loop effective
action was presented as the functional determinant (\ref{G}) of the
special differential operator acting in the analytic subspace of
harmonic superspace. We developed the harmonic superspace
proper-time technique that allows us to calculate such functional
determinants and found the divergent part of the one-loop effective
action in a general form. The result is given by the expression
(\ref{Psi2}) for arbitrary background hypermultiplet $\Omega$.
Taking into account this general result, we calculated the one-loop
divergences in two special cases. First, when the potential-like
functions $L^{++}_{a}$ and $L^{(+4)}$ are absent in action
(\ref{classaction}) and the sigma-model metric $g_{ab}(\Omega)$ is
arbitrary. Second, when the sigma-model metric $g_{ab}$ in
(\ref{classaction}) is flat, the functions $L^{++}_{a}$ and
$L^{(+4)}$ are arbitrary. The background hypermultiplet is still
arbitrary in both cases. It is interesting to note that the one-loop
divergences given by the expression (\ref{Psi2}) are nonlocal in
harmonics but are space-time local. We emphasize that the developed
technique of finding the effective action is manifestly covariant
and preserves the manifest $\cN=2$ supersymmetry at all steps
calculations.

As far as we know, there was only one attempt in literature to
compute the one-loop divergences in some $\cN=2$ supersymmetric
sigma-model \cite{Spence}, which is a special partial case of the
model (\ref{classaction}) considered here. The work \cite{Spence}
approach was based on the construction of divergences of the gauged
$\cN=2$ supersymmetric sigma-model using the divergences of the
$\cN=1$ supersymmetric sigma-model. The result of such an indirect
approach is not manifestly $\cN=2$ supersymmetric and we believe it
requires justification.

As we noted, the calculation of one-loop divergences in this paper
was based on analogies between one-loop effective action for the
general $\cN=2$ model and one-loop effective action for the
hypermultiplet in $\cN=2$ gauge superfield. However, these analogies
are indirect. For example, the structure of the one-loop effective
action \eq{G0} differs from the corresponding effective actions in
$\cN=2$ supersymmetric gauge theories (see, e.g.,
\cite{BBIKO,BBKO,KM1,KM2}) due to the presence of superfield
$\tilde{X}^{(+4)}$. Such a superfield leads to the additional
divergent term in the effective action in comparison with one in SYM
theories.

The approach for calculating the effective action for $\cN=2$
supersymmetric sigma-models developed in the paper is general enough
and can be applied to study the various quantum aspects of the
hypermultiplet theories. In particular, it would be interesting to
apply this general method to calculation of the finite contributions
to the effective action of the model and to study the corresponding
deformation of the initial geometry by quantum corrections. We
believe that the finite contributions can be evaluated by the same
harmonic superfield proper-time method as the divergent ones. We
plan to consider these issues in the forthcoming works.

\appendix
\section{Derivation of the harmonic superspace sigma-model \eq{classaction} from the
general $q^+$--hypermultiplet theory} Let us consider the most
general $\cN=2$ sigma-model with the action
 \bea
 S[q^+] = \int d\zeta^{(-4)} ({\cal L}^{+ A}_a  D^{++} q^{+
 a}_A + {\cal L}^{(+4)})\,,
 \label{genQ}
 \eea
where  $A=1,2$ and $a=1,..,n$. The superfield function ${\cal L}^{+
A}_a (q^+,u)$ and ${\cal L}^{(+4)}(q^+,u)$ are arbitrary function of
the hypermultiplet superfield $q^{+a}_A$ and harmonics $u$. To
derive the action for the $\omega$- hypermultiplet
\eqref{classaction}, we make the change of variables
 \bea
 q^{+a}_A = u_A^+ \omega^a + u^-_A f^{++a}\,,
 \label{defQ}
 \eea
substitute  \eq{defQ} into \eq{genQ} and eliminate the function
$f^{++a}$ using the equation of motion. The decomposition of the
functions ${\cal L}^{+ A}_a (q^+,u)$ and ${\cal L}^{(+4)}(q^+,u)$
over $f^{++a}$ reads
 \bea
 {\cal L}^{+A}_a(u_A^+ \omega^a + u^-_A f^{++a}) &=&  {\cal L}^{+A}_a(u_A^+
 \omega^a) + \frac{ \partial {\cal L}^{+A}_a(u_A^+ \omega^a)}{\partial(u_B^+
\omega^b)} u^-_B  f^{++b}   \nn \\
&&  + \frac{ \partial^2 {\cal L}^{+A}_a(u_A^+
\omega^a)}{\partial(u_B^+ \omega^b)\partial(u_C^+ \omega^c)} u^-_B
f^{++b} \, u^-_C  f^{++c} + \ldots,
  \label{Ldec1} \\  \nn\\
{\cal L}^{(+4)}(u_A^+ \omega^a + u^-_A f^{++a}) &=& {\cal
L}^{(+4)}(u_A^+ \omega^a ) + \frac{\partial {\cal L}^{(+4)}(u_A^+
\omega^a )}{\partial (u_B^+ \omega^b)} u^-_B f^{++b} \nn \\
&& +\frac{\partial^2 {\cal L}^{(+4)}(u_A^+ \omega^a )}{\partial
(u_B^+ \omega^b)\partial (u_C^+ \omega^c)} u^-_B f^{++b}\, u^-_C
f^{++c}+ \dots .
 \label{Ldec2}
 \eea
Substituting the last expressions into \eq{genQ} one obtains
 \bea
 S[\omega, f^{++}] &=& \int d \zeta^{(-4)} \Big( {\cal L}^{+A}_a u^+_A
 D^{++} \omega^a + {\cal L}^{+A}_a  D^{++} (u^-_A f^{++a})  \nn \\
 && +  h_{ab}^{AB} u^+_A u^-_B\,  D^{++} \omega^a f^{++b}
  +  h_{ab}^{AB} u^+_A u^-_B f^{++a} f^{++b} \nn \\
  && +  h_{ab}^{AB} u^+_A u^-_B f^{++b} D^{++}f^{++a}
  + {\cal L}^{(+4)} + h^{(+3)A}_a u^-_A f^{++a} + \dots
 \Big),
 \label{decomp}
 \eea
where we have denoted
 \bea
 h_{ab}^{AB}(\omega) = \frac{ \partial {\cal L}^{+A}_a(u_A^+ \omega^a)}{\partial(u_B^+
\omega^b)} \,,
 \qquad
 h^{(+3) A}_a(\omega) =\frac{\partial {\cal L}^{(+4)}(u_A^+ \omega^a )}{\partial (u_B^+ \omega^b
 )}.
 \label{decomp1}
 \eea
In the expression \eqref{decomp} the ellipsis stands for the higher
derivatives terms of the functions ${\cal L}^{+ A}_a$ and ${\cal
L}^{(+4)}$. To obtain the action under consideration
\eqref{classaction} we  exclude the superfield $f^{++a}$ from
\eqref{decomp} using the corresponding equation of motion. Note that
the action \eq{decomp} contains all powers of the auxiliary
superfield $f^{++a}$. The equation of motion for the superfield
$f^{++a}$ is the non-linear algebraic one, however its can be in
general analyzed. After exclusion of the superfield $f^{++a}$ from
the equation of motion we immediately obtain the the action of the
$\omega$-hypermultiplet model that contains nothing more then the
terms with two $D^{++}$ derivatives of $\omega$, terms with one such
a derivative and terms without derivatives. The coefficients at the
terms with derivative are the functions of $\omega$, the terms
without derivatives also are functions of $\omega$. All the above
functions are expressed through the ${\cal L}^{+ A}_a$ and ${\cal
L}^{(+4)}$ and their derivatives. As a result, we arrive at exactly
the considered harmonic superfield sigma model \eqref{classaction}.

To conclude, in the Appendix we have demonstrated the relation
between the general $q^+$- hypermultiplet theory and harmonic
superfield sigma-model introduced in \eqref{classaction}. Emphasize
that the superfield model \eqref{classaction} is internally
consistent, possesses reparameterization invariance and manifest
$\cN=2$ supersymmetry and has a structure analogous to conventional
sigma-model. It is completely non-contradictory in itself and can be
used for studying the quantum effective action.

\section*{Acknowledgements} The authors are grateful to E.A.
Ivanov and S.M. Kuzenko for valuable comments and discussions and to
S.A. Fedoruk for drawing our attention to the ref. \cite{FIS}. The
work is partially supported by the Ministry of Education of the
Russian Federation, project No. FEWF-2020-003. I.L.B is thankful to
the Laboratory of Theoretical Physics, JINR where the work has been
finalized for hospitality.


\end{document}